\documentclass[twocolumn,showpacs,floatfix,prl]{revtex4}%
\usepackage{graphicx}%
\usepackage{amsmath}%
\usepackage{amsfonts}%
\usepackage{amssymb}
\usepackage{bm}
\usepackage{color}

\def\e{{\epsilon}}
\def\k{{ {\bm k} }}
\def\p{{ {\bm p} }}
\def\q{{ {\bm q} }}
\def\Q{{ {\bm Q} }}
\def\0{{ {\bm 0} }}

\def\w{{\omega}}
\def\a{{\alpha}}

\def\r{{ {\bm r} }}

\allowdisplaybreaks[4]

\begin{document}
\title{
Spin-Fluctuation-Driven Nematic Charge-Density-Wave in Cuprate Superconductors:
Impact of Aslamazov-Larkin-Type Vertex Correction 
}
\author{
Youichi \textsc{Yamakawa}$^{1}$, and
Hiroshi \textsc{Kontani}$^{1}$
}

\date{\today }

\begin{abstract}
We present a microscopic derivation of the nematic charge-density-wave (CDW) 
formation in cuprate superconductors
based on the three-orbital $d$-$p$ Hubbard model,
by introducing the vertex correction (VC) into the charge susceptibility.
The CDW instability at $\q=(\Delta_{\rm FS},0)$, $(0,\Delta_{\rm FS})$ 
appears when the spin fluctuations are strong,
due to the strong charge-spin interference represented by the VC.
Here, $\Delta_{\rm FS}$ is the wavenumber between the neighboring hot spots.
The obtained spin-fluctuation-driven CDW is expressed as the 
``intra-unit-cell orbital order'' accompanied by the charge transfer 
between the neighboring atomic orbitals,
which is actually observed by the STM measurements.
We predict that the cuprate CDW and the nematic orbital order in 
Fe-based superconductors are closely related spin-fluctuation-driven phenomena.

\end{abstract}

\address{
$^1$ Department of Physics, Nagoya University,
Furo-cho, Nagoya 464-8602, Japan. 
}
 

\sloppy

\maketitle


The rich phase diagram of cuprate high-$T_{\rm c}$ superconductors
has been actively studied in condensed matter physics.
The non-Fermi-liquid-like electronic states near the optimally-doped region,
including the $d$-wave transition temperature at $\sim100$ K,
are well understood in terms of the nearly antiferromagnetic 
Fermi liquid picture \cite{Moriya,Yamada-text,Scalapino,ROP},
whereas strong-coupling theories were developed
to describe the under-doped region \cite{PALee}.
In the pseudo-gap state of slightly under-doped cuprates, 
superconducting fluctuations 
play important roles \cite{Yamada-text,ROP,Kivelson,Levin,Sato}.
However, many mysteries concerning in the pseudogap region remain unsolved,
such as the Fermi arc formation 
\cite{Arc-Yoshida,Arc-Kanigel,Arc-Borisenko,Arc-Kondo}
and the small Fermi pockets detected by quantum oscillations
\cite{Doiron-Leyraud}.

The recent discovery of the axial charge-density-wave (CDW) 
parallel to the nearest Cu-Cu direction
in Y-, Bi-, Hg-, and La-based cuprates
by the STM studies
\cite{STM-Hanaguri,STM-Kohsaka,STM-Lawler,STM-Fujita} and
by X-ray scattering studies
\cite{Y-Xray1,Y-Xray2,Y-Xray3,Bi-Xray1,Bi-Xray2,Hg-Xray,La-Xray,p-CDW}
constituted a significant advancement 
in understanding the pseudogap phenomena.
This finding indicates that 
both spin and charge fluctuations cooperatively
develop in under-doped cuprates,
and the interference between charge and spin order parameters 
has been discussed intensively based on various effective 
and microscopic models
\cite{DHLee-PNAS,Kivelson-NJP,Chubu,Sachdev,Metzner,Bulut}.

\begin{figure}[!htb]
\includegraphics[width=.99\linewidth]{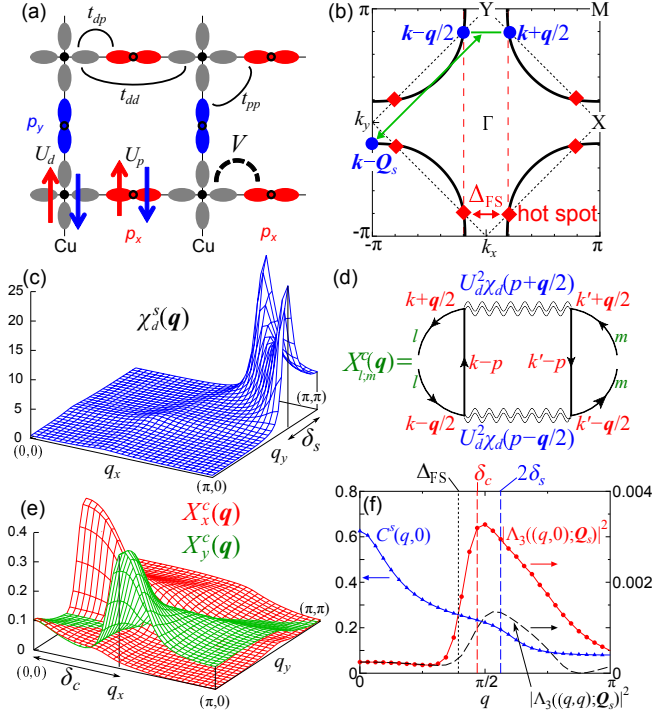}
\caption{(color online)
(a) Three-orbital $d$-$p$ model for the CuO$_2$ plane.
(b) FS for $x=0.1$.
The integrand in Eq. (\ref{eqn:Lambda}) is large when
three points $\k-\q/2$, $\k+\q/2$, $\k-\p$ ($\p=\Q_s$)
are connected by the nesting vectors.
(c) $\chi_d^s(\q)$ given by the RPA. The unit is eV$^{-1}$.
We put $U_d=4.06$ eV and $U_p=0$.
(d) Diagrammatic expression of $X_{m;l}^c(\q)$.
(e) AL-VCs $X_{x}^{c}(\q)$ and $X_{y}^{c}(\q)$.
(f) $C^s(\q)$ and $|\Lambda_3(\q;\Q_s)|^2$ as functions of $\q$.
}
\label{fig:fig1}
\end{figure}

The aim of this paper is to 
present a quantitative microscopic explanation for the 
experimentally observed axial CDW,
since  the diagonal CDW is derived in previous theoretical studies
\cite{Sachdev,Metzner,Bulut}.
Importantly, the CDW wavevector changes with doping,
coinciding with the nesting vector between
the neighboring hot-spots (see Fig. \ref{fig:fig1} (b))
in Y-, Bi- and Hg-based cuprates
\cite{Y-Xray2,Y-Xray3,Bi-Xray1,Bi-Xray2,Hg-Xray}.
In addition, all $p_x$, $p_y$, and $d_{x^2-y^2}$ orbital electrons 
contributes to the CDW formation
\cite{STM-Fujita,p-CDW,Bi-Xray1},
consistently with the local lattice deformation reported in Ref.
\cite{Bianconi}.
The latter fact indicates that {\it $d$-$p$ multiorbital model} 
should be analyzed to reveal the origin of the CDW.
The mean-field-level approximations, such as the 
random-phase-approximation (RPA),
are insufficient to explain these experimental facts.
Thus, we study the role of the vertex correction (VC) in multiorbital models
that describes the strong charge-spin interference 
\cite{Onari-SCVC,Ohno-SCVC,Tsuchiizu,Onari-SCVCS,Tsuchiizu2}.

Other than cuprates,
nematic states are realized in multiorbital systems with strong correlations.
In Fe-pnictides, spin-nematic mechanism \cite{Fernandes}
and orbital-nematic one \cite{OO-FeAs,Onari-SCVC,Onari-SCVCS,Kontani-Raman} 
have been proposed.
In both scenarios,
spin-fluctuation-driven nematicity is discussed.
The latter scenario proposes the orbital-order due to
spin-fluctuation-driven VC, and this scenario is 
applicable even when the spin fluctuations are incommensurate
\cite{Onari-SCVC,Ohno-SCVC,Tsuchiizu,Onari-SCVCS}, like in
Ba(Fe$_{1-x}$Co$_{x}$)$_2$As$_2$ with $x \ge 0.056$ ($T_{\rm N}\le 30$ K).
In cuprates, the VC will develop for both the $d$ and $p$ orbitals,
since both orbitals largely contribute to the 
density-of-states (DOS) at the Fermi level.
Thus, the multiorbital CDW formation in cuprates
could be explained by applying the 
orbital-spin mode-coupling theories
\cite{Onari-SCVC,Onari-SCVCS,Tsuchiizu,Tsuchiizu2,Ohno-SCVC,Kontani-Raman}.

In this paper, we find the significant role of the 
Aslamazov-Larkin VC (AL-VC),
which had not been analyzed in previous studies,
in the formation of the axial CDW in cuprates.
By analyzing the $d$-$p$ Hubbard model with realistic parameters,
we reveal that the axial CDW instability at the wavevectors
$\q=(\Delta_{\rm FS},0)$ and $(0,\Delta_{\rm FS})$,
connected by the neighboring hot-spots, is realized by
the AL-VC in the charge susceptibility.
The CDW emerges only in under-doped region 
since the AL-VC increases in proportion to the spin susceptibility.
The obtained CDW with inter-orbital charge transfer is
consistent with the STM measurements
\cite{STM-Hanaguri,STM-Kohsaka,STM-Lawler,STM-Fujita}.

Figure \ref{fig:fig1} (a) shows the three-orbital $d$-$p$ model 
for cuprates in real space.
The nearest $d$-$p$, $d$-$d$, and $p$-$p$ hopping integrals are
shown as $t_{dp}$, $t_{dd}$, and $t_{pp}$, respectively.
We use the hopping integrals of the first-principles model for La$_2$CuO$_4$ 
listed in Table 2 ($N=0$) of Ref. \cite{Held},
in which the 2nd-nearest ($t_{dp}'$, $t_{pp}'$, $t_{pp}''$)
and the 3rd-nearest ($t_{pp}'''$) hopping integrals exist.
In addition, we include the 3rd-nearest $d$-$d$ hopping $t_{dd}^{\rm 3rd}=-0.1$ eV
to make the Fermi surface (FS) closer to Y- and Bi-based cuprates.
The obtained hole-like FS for the electron filling $n=n_d+n_p=4.9$ 
(hole filling is $x=0.1$) is shown in Fig. \ref{fig:fig1} (b).
We also introduce the on-site Coulomb interactions ($U_d$, $U_p$)
and the nearest $d$-$p$ Coulomb interaction ($V$) 
shown in Fig. \ref{fig:fig1} (a).
The interaction parameters used in the present study is
$(U_d,\ U_p,\ V)\approx(4,\ 0\sim2,\ 0.6)$ in eV 
\cite{parameter-comment}:
The ratios $U_p/U_d$ and $V/U_d$ are consistent with the first principle study 
\cite{model-parameters}.
Later, we will show that
the spin (charge) susceptibility is mainly enlarged by 
$U_d$ ($U_d$ and $V$) sensitively,
whereas both susceptibilities are insensitive to $U_p$
\cite{parameter-comment}.

First, we study the spin and charge susceptibilities
by using the RPA.
We denote $(c_{1}(\k),c_{2}(\k),c_{3}(\k))\equiv
(d_{x^2-y^2}(\k), p_x(\k), p_y(\k))$.
In the RPA without the VC,
the spin (charge) susceptibility in the $3\times3$ matrix form is given as
${\hat \chi}^{s(c)}_{\rm RPA}(q)
={\hat \chi}^{(0)}(q)/\{ {\hat 1}-{\hat \Gamma}^{s(c)}(\q){\hat \chi}^{(0)}(q) \}$,
where ${\hat \Gamma}^{s(c)}(\q)$ is the Coulomb interaction
for the spin (charge) sectors:
${\hat \Gamma}^{s(c)}_{1;1}=(-)U_d$,
${\hat \Gamma}^{s(c)}_{2;2}={\hat \Gamma}^{s(c)}_{3;3}=(-)U_p$, 
${\hat \Gamma}^{c}_{1;2}=-4V\cos(q_x/2)$, and
${\hat \Gamma}^{c}_{1;3}=-4V\cos(q_y/2)$.
$\chi^{(0)}_{l;m}(q)=-T\sum_{k}G_{l,m}(k+q)G_{m,l}(k)$ is the bare bubble, 
and ${\hat G}(k)=(i\e_n+\mu-{\hat H}_\k)^{-1}$.
Here and hereafter, $q\equiv(\q,\w_l)$ and $k\equiv(\k,\e_n)$, where
$\w_l=2l\pi T$ and $\e_n=(2n+1)\pi T$.
Figure \ref{fig:fig1} (c) shows the $d$-orbital spin susceptibility
$\chi_d^s(\q)\equiv \chi^s_{1;1}(\q)$ for $U_d=4.06$ eV and $U_p=0$ 
in the case of $n=4.9$ and $T=0.05$ eV.
(${\hat \chi}^s_{\rm RPA}(\q)$ is independent of $V$.)
The spin Stoner factor $\a_S$, defined as the maximum eigenvalue of
${\hat\Gamma}^s{\hat \chi}^{(0)}(\q)$, is 0.99.

However, the RPA fails to give any CDW instability 
by using the present interaction parameters.
To improve the RPA, we calculate the charge susceptibility 
by including the VC, given as
\begin{eqnarray}
{\hat \chi}^c(q)={\hat \Phi}^c(q)\{{\hat 1}-{\hat \Gamma}^c(\q){\hat \Phi}^c(q)\}^{-1},
 \label{eqn:chiC} 
\end{eqnarray}
where ${\hat \Phi}^c(q)= {\hat \chi}^{(0)}(q)+{\hat X}^c(q)$, and 
${\hat X}^c(q)$ is the irreducible VC for the charge sector.
When the VC is large, 
${\hat \chi}^c(q)$ is enlarged in multiorbital models \cite{Onari-SCVC}.
Here, we consider the AL-VC,
which is the second-order term with respect to the fluctuations,
since it is scaled by the square of the 
spin correlation length $\xi_{\rm AF}^2\sim 1/(1-\a_S)$
in two-dimensional systems
\cite{Onari-SCVC,Miyake}.
The AL-VC gives the nematic orbital order in Fe-pnictides 
\cite{Onari-SCVC,Tsuchiizu}.

The AL-VC increases rapidly with $U_d$
in proportion to $1/(1-\a_S)$ \cite{Onari-SCVC},
whereas it is insensitive to $U_p$ and $V$.
With this in mind, for simplicity,
we present the expression of the AL-VC for $U_p=V=0$:
\begin{eqnarray}
X^c_{l;m}(\q)&=&\frac{TU_d^4}{2}\sum_{p}
\Lambda_l(\q;p)\{  \chi_d^c(p+\frac{\q}{2})\chi_d^c(p-\frac{\q}{2})
\nonumber \\
& &\ \ +3\chi_d^s(p+\frac{\q}{2})\chi_d^s(p-\frac{\q}{2}) \}
\Lambda'_m(\q;p) ,
 \label{eqn:AL} \\
\Lambda_l(\q;p)&=&T\sum_{k}G_{l,1}(k+\frac{\q}{2})G_{1,l}(k-\frac{\q}{2})
G_{1,1}(k-p) ,
\label{eqn:Lambda} 
\end{eqnarray}
where $p=(\p,\w_m)$, $\chi_d^{c(s)}(q)\equiv\chi_{1;1}^{c(s)}(q)$,
and $\Lambda'_m(\q;p)\equiv \Lambda_m(-\q;p)+\Lambda_m(-\q;-p)$:
The relation $\Lambda'_m(\q;p)=2\Lambda_m(\q;p)$ holds in the present model.
Its diagrammatic expression is shown in Fig. \ref{fig:fig1} (d).
The dominant contribution of the AL-VC 
has been verified by the functional RG method \cite{Tsuchiizu,Tsuchiizu2}.
In the SC-VC method \cite{Onari-SCVC},
we calculate both ${\hat \chi}^{c,s}$ and ${\hat X}^{c,s}$ self-consistently.
In the present model, however, we verified that 
the positive feedback effect from ${\hat \chi}^c$ to ${\hat X}^c$, 
which is important in Fe-pnictides \cite{Onari-SCVC},
is very small.
Thus, we can safely replace ${\hat \chi}^{c,s}$ in Eq. (\ref{eqn:AL}) with 
${\hat \chi}^{c,s}_{\rm RPA}$.
We verified that the Maki-Thompson (MT) VC
is considerably smaller than the AL-VC;
see Refs \cite{Ohno-SCVC,Tsuchiizu} and Supplemental Material \cite{suppl}.

In cuprates, 
both $d$ orbital and $p$ orbital AL-VCs are strongly enhanced
when $\chi^s_{1;1}(\q)$ is large, since
the $p$-orbital DOS is large at the Fermi level \cite{comment1}.
Figure \ref{fig:fig1} (e) shows the obtained $X^c_x(\q)\equiv X^c_{2;2}(\q)$ 
and $X^c_y(\q)\equiv X^c_{3;3}(\q)$ 
for the parameters used in Fig. \ref{fig:fig1} (c).
$X^c_{y(x)}(\q)$ shows the maximum at $\Q_c=(\delta_c,0)$ ($\Q_c'=(0,\delta_c)$),
and it is about one-third of $X^c_d(\q)\equiv X^c_{1;1}(\q)$ in magnitude.
The AL-VC for $p_y$-orbital is approximately given as
\begin{eqnarray}
X^c_y(\q) &\sim& U_d^4|\Lambda_3(\q;\Q_s)|^2C^s(\q) ,
 \label{eqn:X-ap} \\
C^s(\q)&=& T\sum_{p}\chi^s_d(p+\q/2)\chi^s_d(p-\q/2),
\end{eqnarray}
where $\Q_s=(\pi,\pi)$.
The $\q$-dependences of these functions along the $q_x$-axis
are shown in Fig. \ref{fig:fig1} (f).
Here, $C^s(\q)$ has a maximum at $\q=0$, and its width 
is about $2\delta_s$:
A weak shoulder structure of $C^s(\q)$ at $\q=(2\delta_s,0)$
reflects the incommensurate peaks of $\chi^s(\q)$ at $\q=(\pi\pm\delta_s,\pi)$.
On the other hand,
$|\Lambda_3(\q;\Q_s)|^2$ in Fig. \ref{fig:fig1} (f) takes the maximum value 
at $\q\approx(\Delta_{\rm FS},0)$, reflecting the nesting 
between the hot spots.
In fact, the integrand of Eq. (\ref{eqn:Lambda}) 
is large in magnitude when $\k+\q/2$, $\k-\q/2$, $\k-\Q_s$ are on the FS and
connected by the nesting vector.
($p_{y(x)}$ orbital has large weight around Y (X) point.)
Thus, the large peak of $X_y^c(\q)$ at $\q=(\delta_c,0)$
originates from the strong $\q$-dependence of 
the three-point vertex $\Lambda_3(\q;\Q_s)$
in the present parameters.

Note that
$|\Lambda_3(\q;\Q_s)|^2$ at $\q=(\delta_c,\delta_c)$ is much smaller than 
that at $\q=(\delta_c,0)$
as shown in Fig. \ref{fig:fig1} (f).
Thus, the axial CDW is selected by the 
strong $\q$-dependence of $|\Lambda_3(\q;\Q_s)|^2$,
contrary to many previous theoretical studies that predicted the diagonal CDW
\cite{Sachdev,Metzner,Bulut}.

\begin{figure}[!htb]
\includegraphics[width=.99\linewidth]{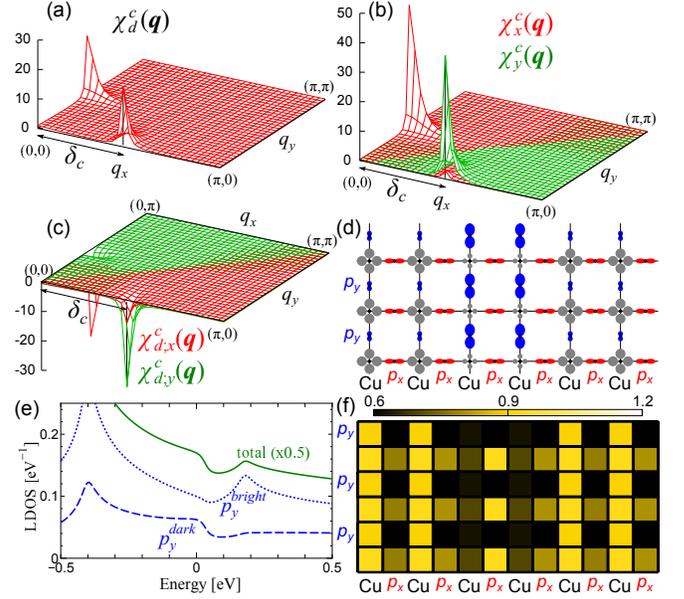}
\caption{(color online)
Charge susceptibilities with VC:
(a) $\chi^c_{d}(\q)$, (b)  $\chi^c_{x}(\q)$ and $\chi^c_{y}(\q)$,
(c) $\chi^c_{d;x}(\q)$ and $\chi^c_{d;y}(\q)$.
We put $U_d=4.06$ eV, $U_p=0$, and $V=0.65$ eV.
(d) A possible charge pattern of the CDW ($\delta_c=\pi/2$).
Since the charge transfer between the 
neighboring $n_{y}$ and $n_d$ occurs,  
($n_{y}$, $n_d$) are in antiphase in the intra-unit-cell.
(e) Total DOS and local DOSs at two $p_y$-sites in the CDW state with 
$\Q_c=(\pi/2,0)$.
(f) $R(\r,E)$ in the CDW state for $E=0.2$ eV.
}
\label{fig:fig2}
\end{figure}

Thanks to large AL-VC,
the charge susceptibility in Eq. (\ref{eqn:chiC})
are enhanced at $\q=(\delta_c,0)$, and it diverges when
the charge Stoner factor $\a_C$, defined as the maximum eigenvalue of
${\hat\Gamma}^c(\q){\hat \Phi}^c(\q)$, reaches unity.
The CDW ($\a_C=1$) is realized due to the finite off-diagonal 
elements of ${\hat \Gamma}^c$; $\Gamma_{1;2(3)}^c\propto V$.
We will show later that the CDW emerges when $X^c_y(\Q_c)\gtrsim U_d/16V^2$,
and $X^c_y(\Q_c)$ scales as $(1-\a_S)^{-1}$.
Thus, the larger $V$ is, the smaller $\a_S$ for realizing the CDW is.
In Figs. \ref{fig:fig2} (a)-(c), we show the largest three 
susceptibilities, $\chi^c_{d}(\q)\equiv \chi^c_{1;1}(\q)$, 
$\chi^c_{x(y)}(\q)\equiv \chi^c_{2;2(3;3)}(\q)$, and
$\chi^c_{d;x(d;y)}(\q)\equiv \chi^c_{1;2(1;3)}(\q)$ at $n=4.9$,
in the case of $U_d=4.06$ eV and $U_p=0$ ($\a_S=0.99$).
We also put $V=0.65$ eV, at which $\a_C$ reaches $0.99$.
Both $\chi^c_{d}$ and $\chi^c_{y}$ show large positive values 
at $\q=\Q_c$, whereas  $\chi^c_{d;y}$ develops negatively.
The charge density modulation 
($\Delta n_d(\q), \Delta n_x(\q), \Delta n_y(\q)$)
at $\q=(\delta_c,0)$ is proportional to the form factor
that is given by the eigenvector of ${\hat \chi}^c(\q)$
for the largest eigenvalue.
The form factor for Figs. \ref{fig:fig2} (a)-(c)
is given as ${\bm f}=(-0.56, 0.21, 0.80)$, which means that 
($d$, $p_y$) orbitals form the ``antiphase CDW state''.
(Note that the form factor is sensitive to the model parameters.)
A possible charge distribution patterns
for $\Q_c=(\pi/2,0)$ is depicted in Fig. \ref{fig:fig2} (d).
We verified that the antiphase CDW with respect to the nearest ($n_x$, $n_y$)
develops if we introduce small repulsion $V_{p_x p_y}$;
see Supplemental Material \cite{suppl}.

\begin{figure}[!htb]
\includegraphics[width=.90\linewidth]{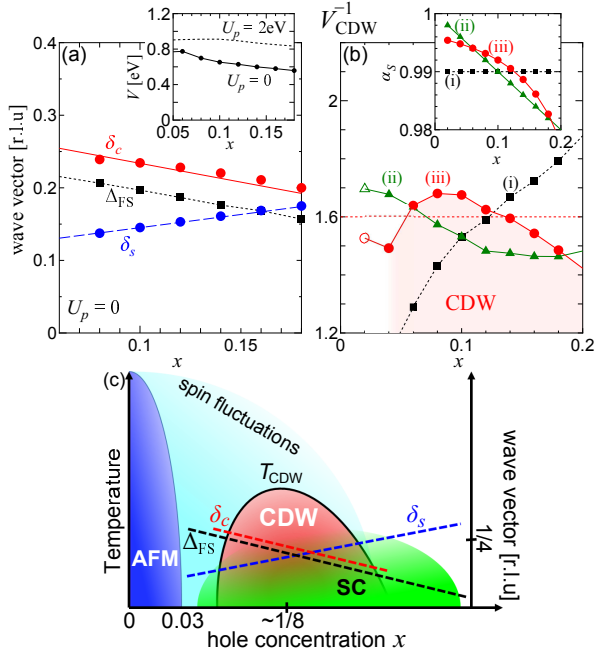}
\caption{(color online)
(a) $\Delta_{\rm FS}$, $\delta_s$ and $\delta_c$
obtained in the $d$-$p$ model as functions of $x=5-n$.
(Inset) The values of $V$ required to give the CDW. 
(b) The values of $V_{\rm CDW}^{-1}$ as a function of $x$,
which will depict a qualitative behavior of $T_{\rm CDW}$.
Note that $\Q_c$ shifts to ${\bm 0}$ for $x\le0.02$.
(Inset) $\a_S(x)$ (i)-(iii) are shown.
(c) Schematic phase diagram of cuprates.
}
\label{fig:fig3}
\end{figure}

Here, we calculate the $d$- and $p$-orbital local DOSs 
in the nematic CDW shown in Fig. \ref{fig:fig2} (d),
under the CDW order parameter at $\bm r$ predicted by the present theory 
$(\Delta n_d, \Delta n_x, \Delta n_y)
={\bm f}\cdot b\cos((\pi/2)(r_x+1/2))$.
Figure \ref{fig:fig2} (e) shows the obtained local DOS $N(\r,\e)$
at two $p_y$-sites, and the total DOS for $b\sim 0.08$.
The pseudo-gap appears due to the CDW hybridization gap.
(Here, we put $n=5.0$ since $\delta_c=\pi/2$ is achieved at $x=0$
in the present single-layer model; see Fig. \ref{fig:fig3} (a).
This will be justified in double-layer YBCO and BSCCO since
the FS of the bonding-band is large.)
In Fig. \ref{fig:fig2} (f), we show the obtained ratio
$R(\r,E)=\int_0^{E}N(\r,\e)d\e/\int_{-E}^0N(\r,\e)d\e$
at Cu and O sites for $E=0.2$ eV.
The realized intra-unit-cell nematic order looks
similar to the recent STM results
\cite{STM-Hanaguri,STM-Kohsaka,STM-Fujita}.
Moreover, the Fermi arc structure found by ARPES
\cite{Arc-Kanigel,Arc-Borisenko,Arc-Kondo,Arc-Yoshida}
would be formed by the single-Q or double-Q CDW order
\cite{doubleQ,smectic-theory}.
The Fermi arc structure similar to cuprates
was recently reported in Sr$_2$IrO$_4$
\cite{Kim}.

Here, we present an analytic explanation why the nematic CDW 
is realized by the AL-VC in the presence of small $V$.
To simplify the discussion,
we consider only $\Phi^c_{d}\equiv \Phi^c_{1;1}$ and 
$\Phi^c_{y}\equiv \Phi^c_{3;3}$ and put $U_p=0$ in Eq. (\ref{eqn:chiC}).
The obtained results at $\q\approx\Q_c$ are
\begin{eqnarray}
\chi^c_{d}(\q)&=& \Phi^c_{d}(\q)/D(\q),
\\
\chi^c_{y}(\q)&=&\Phi^c_{y}(\q)\{ 1+ U_d\Phi^c_{d}(\q) \}/D(\q) ,
\end{eqnarray}
and $\chi^c_{d;y}(\q)= -4V \Phi^c_{y}(\q)\cdot\chi^c_{d}(\q)$, where 
$D(\q)=1+\Phi^c_{d}(\q)\{ U_d-16V^2 \Phi^c_{y}(\q) \}$.
Thus, the charge susceptibilities develop divergently when
$\Phi^c_{y}(\Q_c)$ is greater than $U_d/16V^2$ due to the AL-VC.
Note that $U_d/16V^2 \ll 1$ according to
the first principle study \cite{model-parameters}.

In the RPA without the VC,
${\hat \chi}^c(\q)$ diverges when $V$ is larger than $2.7$ eV,
which is much larger than the first principle value \cite{model-parameters}.
Worse still, the divergence occurs at $\q={\bm 0}$ in this model.
Thus, the VC is indispensable to realizing the stripe CDW state.
The RPA analysis on a $d$-$p$ model with $V_{p_x p_y}$ was 
done in Ref. \cite{Bulut} in detail.

Now, we study the hole carrier ($x\equiv5-n$) dependence of the CDW state.
Figure \ref{fig:fig3} (a) shows $\Delta_{\rm FS}$, and obtained 
$\delta_s$ and $\delta_c$ for $U_p=0$,
by choosing $U_d$ and $V$ so as to satisfy $\a_S=\a_C=0.99$.
The inset of (a) shows the used $V$,
which is much smaller than the first principle value
for both $U_p=0$ and $2$ eV.
Also, the used $U_d$ is $4.0\sim4.1$ eV.
Here, $\delta_s$ decreases for $x\rightarrow0$
as observed by neutron measurements.
In contrast, $\delta_c$ increases as $x\rightarrow0$
with satisfying the relation $\delta_c\approx \Delta_{\rm FS}$,
which is widely observed in Y-, Bi- and Hg-based compounds 
\cite{Y-Xray2,Y-Xray3,Bi-Xray1,Bi-Xray2,Hg-Xray}.
Also, the relation $\delta_c \gtrsim \Delta_{\rm FS}$ is
consistently with experiments.

Here, we explain why the CDW appears only in slightly under-doped region.
In Fig. \ref{fig:fig3} (b), we show 
the inverse of $V$ at the CDW boundary, $V_{\rm CDW}$, for $U_p=0$,
by adjusting $U_d$ to satisfy $\a_S=\a_S(x)$.
In the case of (i) $\a_S(x)=0.99$,
$V_{\rm CDW}^{-1}$ decreases as $x\rightarrow0$,
since the AL-VC at $\q=(\Delta_{\rm FS},0)$,
which is proportional to $C^s(\Delta_{\rm FS},0)$ in Fig. \ref{fig:fig1} (f),
becomes small when $\Delta_{\rm FS}\gg\delta_s$.
However, $\a_S$ decreases with $x$ in cuprates, which
is reproduced by the FLEX approximation using a fixed $U_d$ \cite{ROP}.
Thus, the CDW should disappear in over-doped region
since the AL-VC is scaled by $\xi_{\rm AF}^2\propto 1/(1-\a_S)$
\cite{Onari-SCVC}.
For this reason, 
we also set $1/\sqrt{1-\a_S}$ as
(ii) $3.3/\sqrt{x}$ and (iii)  $16(1-2.9x)$.
In case (iii), if we fix 
$V^{-1}=1.6$ (dotted line),
the CDW is realized only for $0.06<x<0.12$.
Thus, the phase diagram in Fig. \ref{fig:fig3} (c)
is well understood. 

\begin{figure}[!htb]
\includegraphics[width=.9\linewidth]{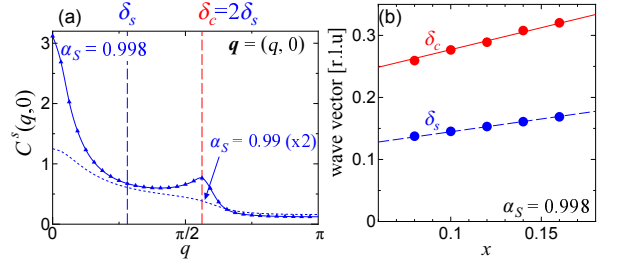}
\caption{(color online)
(a) $C^s(\q)$ for $x=0.1$ and 
(b) $\delta_s$ and $\delta_c$ in the case of $\a_S=0.998$ ($U_d=4.09$ eV)
and $V_{\rm CDW}\approx0.35$eV.}
\label{fig:fig4}
\end{figure}

In La-based compounds, the relation $\delta_c\approx 2\delta_s$ 
is satisfied \cite{La-Xray}, differently from other compounds.
To understand this fact in the present theory,
we study the case $\a_S=0.998$,
in which $\chi^s(\Q_s')$ at $\Q_s'=(\pi\pm\delta_s,\pi)$ reaches $100$ eV$^{-1}$,
which is still smaller than 
the neutron experimental data in 60K YBCO
\cite{neutron-YBCO}.
In this case, the incommensurate peak in $\chi^s(\q)$ becomes sharper
as observed in La-based compounds.
Then, the shoulder peak in $C^s(\q)$ at 
$\q=2\Q_s'=(2\delta_s,0)$ 
becomes prominent as shown in Fig. \ref{fig:fig4} (a).
For this reason, the CDW wavevector $\delta_c$ is fixed at $2\delta_s$
as shown in Fig. \ref{fig:fig4} (b).
In this case, $V_{\rm CDW}\approx 0.35$ eV.
Thus, the relation $\delta_c\approx 2\delta_s$ can be realized 
when $\chi^s(\q)$ shows clear incommensurate peak structure.
Therefore, the present CDW mechanism due to AL-VC can explain 
both the relations $\delta_c \sim \Delta_{\rm FS}$ and $\delta_c \sim 2\delta_s$,
and the latter is realized when $\chi^s(\q)$ shows clear incommensurate peaks.
This result would be a great hint to understand the CDW in LSCO.
Note that the relation $\Q_c=2\Q_s'$ is naturally understood 
since the AL-VC represents the interference of two magnons.

In our theory, the CDW originates from the repulsive interactions, and 
the $e$-ph interaction is unnecessary.
In real compounds, the Coulomb-interaction-driven CDW fluctuations 
couple to the lattice due to finite $e$-ph interactions, so
the Kohn anomaly will emerge.
\cite{Andersen,Reznik,Dev}.

Finally, we discuss the close relation between the CDW in cuprates
and the nematic orbital order in Fe-pnictides.
In both systems, the charge-spin interference,
which is given in the AL-VC,
causes the inter-orbital charge transfer when $\xi_{\rm AF}\gg1$
\cite{Onari-SCVC,Onari-SCVCS}.
In Fe-pnictides, both $\q={\bm 0}$ and $\q\ne{\bm 0}$ $d$-orbital 
orders/fluctuations have been discussed intensively
\cite{DHLee-PNAS,Onari-SCVC,Onari-SCVCS,Kasahara},
and both fluctuations will contribute to the superconductivity.

In summary,
we revealed that the axial nematic CDW in under-doped cuprates 
originates from the AL-VCs in ${\hat \Phi}^c(q)$, which describes the 
interference of two-magnons.
It is shown that 
both the spin fluctuations at $\Q_s\approx(\pi,\pi)$
and charge-orbital fluctuations 
at $\Q_c\approx(\Delta_{\rm FS},0), (0,\Delta_{\rm FS})$ develop mutually.
(This VC-driven CDW cannot emerge in the single-orbital Hubbard model, 
as we discuss in Supplemental Material \cite{suppl}.)
We predict that 
charge-orbital-spin multimode fluctuations
emerge ubiquitously in cuprates, Fe-pnictides, and other 
strongly correlated electron systems,
due to the significant contribution of the AL-VC.

\acknowledgements
We are grateful to S. Onari, M. Tsuchiizu,
S. Borisenko, Y. Matsuda and T. Hanaguri
for fruitful discussions.
This study has been supported by Grants-in-Aid for Scientific 
Research from MEXT of Japan.



\newpage

\section{
[Supplemental Material]
}

In the main text,
we investigate the $d$-$p$ Hubbard model with 
repulsive Coulomb interactions.
The charge-spin interference due to Aslamazov-Larkin (AL) type 
vertex correction (VC) is important in the presence of strong spin fluctuations.
For this reason, the CDW instability at wavevectors
$\q=(\Delta_{\rm FS},0)$ and $(0,\Delta_{\rm FS})$,
connected by the neighboring hot-spots, is promoted by the VC.

\section{CDW in One-Orbital Hubbard Model with $e$-ph Interaction}
Here, we study the CDW formation in the single $d$-orbital Hubbard model.
As far as only repulsive Coulomb interaction is taken into account,
the VC driven CDW cannot emerge in the single-orbital Hubbard model.
However, we show that the CDW is formed due to the cooperation of
the VC and the $e$-ph interaction.

\begin{figure}[!htb]
\includegraphics[width=.99\linewidth]{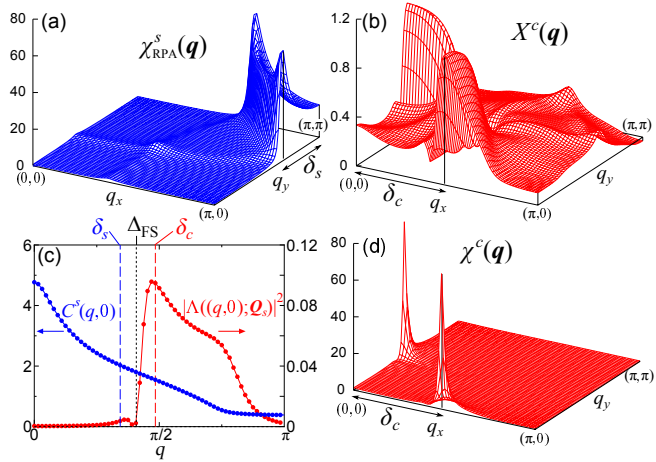}
\caption{(color online)
(a) $\chi^s(\q)$ in one-orbital Hubbard model in the RPA for $\a_S=0.99$.
(b) $X^c(\q)$, and (c) $C^s(\q)$ and $|\Lambda(\q,\Q_s)|^2$.
(d) $\chi^c(\q)$ given by the SC-VC method with the $e$-ph interaction.
}
\label{fig:figs1}
\end{figure}

The band-dispersion is 
$\e_\k=2t(\cos k_x+\cos k_y)+4t'\cos k_x\cos k_y+2t''(\cos 2k_x+\cos 2k_y)$,
where $t=-0.5$ eV, $t'/t=-1/6$ and $t''/t=1/5$ for YBCO \cite{ROP}.
In the random-phase-approximation (RPA) without the VC,
the spin (charge) susceptibility is given as
$\chi^{s(c)}_{\rm RPA}(q)=\chi^{(0)}(q)/\{ 1-(+)U\chi^{(0)}(q) \}$:
$U$ is the on-site Coulomb interaction,
$\chi^{(0)}(q)=-T\sum_{k}G(k+q)G(k)$ is the bare bubble, and
$G(k)=(i\e_n+\mu-\e_\k)^{-1}$.
Here and hereafter, $q\equiv(\q,\w_l)$ and $k\equiv(\k,\e_n)$, where
$\w_l=2l\pi T$ and $\e_n=(2n+1)\pi T$.
Figure \ref{fig:figs1} (a) shows 
$\chi^{s}_{\rm RPA}(\q)\equiv \chi^{s}_{\rm RPA}(\q,0)$ for $U=1.65$ eV, 
$x=0.1$ and $T=0.025$ eV.
The spin Stoner factor $\a_S\equiv\max_\q\{U\chi^{(0)}(\q)\}$ is 0.99.
In contrast, $\chi^{c}(\q)$ is suppressed by $U$ within the RPA.

Next, we discuss the charge susceptibility beyond the RPA
by taking the VC into account.
It is given as
\begin{eqnarray}
\chi^c(q)&=&\Phi^c(q)/\{ 1+U\Phi^c(q) \},
 \label{eqn:chiC-one} 
\end{eqnarray}
where $\Phi^c(q)= \chi^{(0)}(q)+X^c(q)$, and
$X^c(q)$ is the AL-type VC \cite{Onari-SCVC,Onari-SCVCS}.
At $\w_l=0$, the AL-term is given as
\begin{eqnarray}
X^c(\q)&=&TU^4\sum_{p}
\Lambda(\q;p)\{  {\chi}^c(p+\q/2){\chi}^c(p-\q/2)
\nonumber \\
& &\ \ +3{\chi}^s(p+\q/2){\chi}^s(p-\q/2) \}
\Lambda(\q;p) ,
 \label{eqn:AL-2} \\
\Lambda(\q;p)&=&T\sum_{k}G(k+\q/2)G(k-\q/2)G(k-p) ,
\label{eqn:Lambda-2} 
\end{eqnarray}
where $p=(\p,\w_m)$.
Figure \ref{fig:figs1} (b) shows the obtained $X^c(\q)$ 
for $U=1.65$ eV ($\a_S=0.99$),
which shows the maximum at $\Q_c=(\delta_c,0)$ and $\Q_c'=(0,\delta_c)$,
and the relation $\delta_c\approx\Delta_{\rm FS}$ is satisfied.
The CDW instability will be caused by the large $X^c(\Q_c)$,
which is much larger than ${\rm max}_\q\{\chi^{(0)}(\q)\}\approx0.6$.
Here, the $\q$-dependence of $X^c(\q)$ is mainly given by
the three-point vertex.
To show this, 
we approximate the AL-type VC for $\a_S\lesssim1$ as
\begin{eqnarray}
X^c(\q) \sim U^4|\Lambda(\q;\Q_s)|^2C^s(\q) ,
\end{eqnarray}
where $\Q_s=(\pi,\pi)$ and 
$C^s(\q)\equiv T\sum_{p}\chi^s_{\rm RPA}(p+\q/2)\chi^s_{\rm RPA}(p-\q/2)$.
Figure \ref{fig:figs1} (c) shows the $\q$-dependences of
$|\Lambda(\q;\Q_s)|^2$ and $C^s(\q)$ along the $q_x$-axis.
It is apparent that the large peak of $X^c(\q)$ at $\q=(\delta_c,0)$
originates from the three-point vertex.

Although the charge VC becomes very large 
in the presence of strong spin fluctuations,
$\chi^c(\q)$ in Eq. (\ref{eqn:chiC-one}) cannot exceed $1/U$.
In real compounds, however, strong attractive interaction $-g(\q)$ 
due to the buckling mode with $\q\sim\Q_c$ 
had been predicted by the first principle study \cite{Andersen}.
In this case, $U$ in Eq. (\ref{eqn:chiC-one}) is replaced with $U-2g(\q)$,
and then $\chi^c(\Q_c)$ can be strongly enlarged if $U-2g(\Q_c)$ is negative.
That is, {\it the CDW instability due to the $e$-ph interaction 
is effectively enlarged by the spin-fluctuation driven AL-type VC.}
In Fig. \ref{fig:figs1} (d),
we show the strong developed $\chi^c(\q)$ obtained for 
$U=1.65$ eV and $g(\q)=1.16$ eV.

In the main text, we show that the CDW due to the VC
is much easily realized in the three-orbital $d$-$p$ model
with degenerate $p_x$ and $p_y$ orbitals,
{\it without introducing the $e$-ph interactions}.

\section{CDW in Three-Orbital Hubbard Model with $V_{p_x,p_y}$}

In the main text, we studied the three-orbital model 
with on-site Coulomb interactions ($U_d$, $U_p$) and 
inter-site Coulomb interaction $V$ between the nearest Cu-O sites.
Here, we introduce the Coulomb interaction between the nearest O-sites,
$V_{p_x,p_y}$, and discuss the VC-driven CDW.

\begin{figure}[!htb]
\includegraphics[width=.99\linewidth]{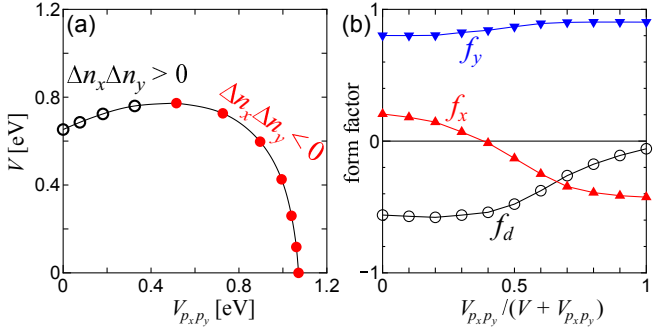}
\caption{(color online)
(a) The CDW boundary in the $V$-$V_{p_x,p_y}$ plane at $n=4.9$,
in the case of $\a_S=0.99$.
(b) The form factor on the CDW boundary as function of $V_{p_x,p_y}/V$.
}
\label{fig:figs2}
\end{figure}

Figure \ref{fig:figs2} (a) shows the
CDW boundary ($\a_C=1$) in the $V$-$V_{p_x,p_y}$ plane at $n=4.9$,
in the case of $\a_S=0.99$ ($U_d\sim4.06$ eV).
Here, $V_{p_x,p_y}=0$ corresponds to the main text.
Thus, the nematic CDW at $\q=\Q_c$ is realized
in the presence of small $V$ and/or $V_{p_x,p_y}$,
by taking the AL-term into account.

Figure \ref{fig:figs2} (b) shows the 
form factor on the CDW boundary, ${\bm f}=(f_d,f_x,f_y)$, which is given by 
the eigenvector of ${\hat \chi}^c(\q)$ for the largest eigenvalue.
The charge density modulation ($\Delta n_d, \Delta n_x, \Delta n_y$)
at $\q=(\delta_c,0)$ is proportional to ${\bm f}$.
We see that the form factor sensitively depends on the ratio $V_{p_x,p_y}/V$.
We stress that 
the CDW order with $\Delta n_x\cdot\Delta n_y<0$ is realized
when $V_{p_x,p_y}$ is comparable or larger than $V/2$.
Recently, the CDW form factors are determined in several compounds
by using the resonant $X$-ray measurements.
By comparing the experimental CDW form factors 
and Fig. \ref{fig:figs2} (b), we can determine the ratio $V_{p_x,p_y}/V$
in each real compound.

The CDW order parameter $\Delta n_l(\r)=\langle n_l(\r)\rangle - n_l^0$
($l=1,2,3$) is given by
\begin{eqnarray}
\Delta n_l(\r) = -T \sum_{n} [G_{l}(\r,\e_n;\{n_l\})-G_{l}(\r,\e_n; \{n_l^0\})]
e^{i\e_n\cdot 0} ,
\end{eqnarray}
where $G_l$ is the Green function.
By expanding the Green function with respect to $\Delta n_l(\r)$, we obtain
\begin{eqnarray}
\Delta n_l(\q) = \sum_{m,m'} \Phi_{l,m}^c(\q)\Gamma^c_{m,m'}(\q)
\Delta n_{m'}(\q) + O((\Delta n_l)^3),
\end{eqnarray}
where ${\hat \Phi}^c(\q)= {\hat \chi}^0(\q)+{\hat X}^c(\q)$:
Here, ${\hat \chi}^0(\q)$ is the bare susceptibility,
and ${\hat X}^c(\q)$ is the charge VC beyond the mean-field approximation,
given by the Ward identity with respect to the self-energy.
${\hat \Gamma}^c(\q)$ is the bare Coulomb interaction for the 
charge sector introduced in the main text.
Then, the linearized equation with respect to $\Delta n_l(\q)$ is
\begin{eqnarray}
\lambda \Delta n_l(\q) = \sum_{m,m'} \Phi_{l,m}^c(\q)\Gamma^c_{m,m'}(\q)
\Delta n_{m'}(\q) ,
\end{eqnarray}
where the eigenvalue $\lambda$ reaches unity at $T=T_{\rm CDW}$.
Therefore, the form factor 
${\bm f}(\q)\propto \Delta {\bm n}(\q)$ at $T=T_{\rm CDW}$
is given by the eigenvector of ${\hat \Phi}^c(\q){\hat \Gamma}^c(\q)$ 
for the largest $\lambda=\a_C$, which is the charge Stoner factor.
When $\a_C\rightarrow 1$,
it is easy to show that the form factor ${\bm f}(\q)$ is approximately 
equal to the eigenvector of the charge susceptibility 
${\hat \chi}^c(\q)=[{\hat 1}-{\hat \Phi}^c(\q){\hat \Gamma}^c(\q)]^{-1}{\hat \Phi}^c(\q)$
for the largest eigenvalue.
In other words,  under the condition $|{\bm f}|=|{\bm f}'|=1$,
the maximum of the inner product 
$({\bm f}',{\hat \chi}^c{\bm f}')$ is realized for 
${\bm f}'\approx {\bm f}$ in the case of $\a_C\approx 1$,
if the largest eigenvalue of 
${\hat \Phi}^c(\q){\hat \Gamma}^c(\q)$ is nondegenerate.

\section{Comparison between AL-VC and MT-VC}

\begin{figure}[!htb]
\includegraphics[width=.99\linewidth]{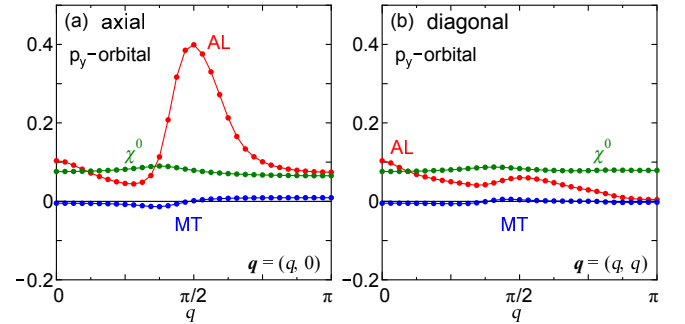}
\caption{(color online)
$\q$-dependences of AL-VC and MT-VC for $p_y$-orbital,
$X_{3;3}^{\rm AL}(\q)$ and $X_{3;3}^{\rm MT}(\q)$ respectively.
The bare susceptibility for $p_y$-orbital, $\chi^{(0)}_{3;3}(\q)$,
is also shown.
}
\label{fig:figS3}
\end{figure}

In the main text, 
we studied the role of the VC for the CDW formation in cuprates,
by taking only the AL-VC into account.
Here, we show the $\q$-dependence of the AL-VC 
as well as that of the Maki-Thompson (MT) VC:
The MT-VC is the first-order term with respect to the fluctuations.
Figure \ref{fig:figS3} shows the AL-VC and MT-VC for $p_y$-orbital,
$X_{3,3}^{\rm AL}(\q)$ and $X_{3,3}^{\rm MT}(\q)$ respectively,
in the three-orbital $d$-$p$ Hubbard model studied in the main text.
Here, we put $U_d=4.06$ eV, and $U_p=V=0$ ($\a_S=0.99$).

Thus, it is verified that the MT-VC
is considerably smaller than the AL-VC in the present $d$-$p$ Hubbard model,
similarly to other multiorbital models studied previously
\cite{Onari-SCVC,Onari-SCVCS}.


\begin{thebibliography}{99}

\bibitem{Moriya}
T. Moriya and K. Ueda, Adv. Phys. {\bf 49}, 555 (2000).

\bibitem{Yamada-text}
K. Yamada: {\it Electron Correlation in Metals} 
(Cambridge Univ. Press 2004).

\bibitem{Scalapino}
D. J. Scalapino, Phys. Rep. {\bf 250}, 329 (1995).

\bibitem{ROP}
H. Kontani, Rep. Prog. Phys. {\bf 71}, 026501 (2008);
H. Kontani, {\it Transport Phenomena in Strongly Correlated Fermi Liquids}
(Springer-Verlag Berlin and Heidelberg GmbH \& Co. K, 2013).

\bibitem{PALee}
P. A. Lee, N. Nagaosa, and X.-G. Wen, Rev. Mod. Phys. {\bf 78}, 17 (2006);
M. Ogata and H. Fukuyama, 
Rep. Prog. Phys. {\bf 71}, 036501 (2008).

\bibitem{Kivelson}
V. J. Emery and S. A. Kivelson, Nature {\bf 374}, 434 (1995). 

\bibitem{Levin}
Q. Chen, I. Kosztin, B. Jank\'{o}, and K. Levin, 
Phys. Rev. Lett. {\bf 81}, 4708 (1998).

\bibitem{Sato}
M. Sato, H. Harashina, J. Takeda, S. Yoshii, Y. Kobayashi, and K. Kakurai,
J. Phys. Chem. Solids {\bf 62}, 7 (2001).

\bibitem{Arc-Yoshida}
T. Yoshida {\it et al}.,
Phys. Rev. Lett. {\bf 91}, 027001 (2003);
T. Yoshida {\it et al}.,
J. Phys. Soc. Jpn. {\bf 81}, 011006 (2012).

\bibitem{Arc-Kanigel}
A. Kanigel {\it et al}.,
Nature Physics {\bf 2}, 447 (2006).

\bibitem{Arc-Borisenko}
A. A. Kordyuk, S. V. Borisenko, V. B. Zabolotnyy, R. Schuster, D. S. Inosov, 
D. V. Evtushinsky, A. I. Plyushchay, R. Follath, A. Varykhalov, L. Patthey, 
and H. Berger, Phys. Rev. B {\bf 79}, 020504(R) (2009).

\bibitem{Arc-Kondo}
T. Kondo et al., 
Nature Physics {\bf 7}, 21 (2011).

\bibitem{Doiron-Leyraud}
N. Doiron-Leyraud, C. Proust, D. LeBoeuf, J. Levallois, J.-B. Bonnemaison, 
R. Liang, D. A. Bonn, W. N. Hardy, and L. Taillefer,
Nature {\bf 447}, 565 (2007).

\bibitem{STM-Hanaguri}
T. Hanaguri, C. Lupien, Y. Kohsaka, D.-H. Lee, M. Azuma, M. Takano, 
H. Takagi, and J. C. Davis, Nature {\bf 430}, 1001 (2004).

\bibitem{STM-Kohsaka}
Y. Kohsaka, T. Hanaguri, M. Azuma, M. Takano, J. C. Davis, and H. Takagi,
Nature Physics {\bf 8}, 534 (2012). 

\bibitem{STM-Lawler}
M. J. Lawler {\it et al}.,
Nature {\bf 466}, 347 (2010).

\bibitem{STM-Fujita}
K. Fujita {\it et al}.,
Proc. Natl. Acad. Sci. USA, {\bf 110}, E3026 (2014).

\bibitem{Y-Xray1}
G. Ghiringhelli {\it et al}.,
Science {\bf 337}, 821 (2012).

\bibitem{Y-Xray2}
J. Chang {\it et al}.,
Nature Physics {\bf 8}, 871 (2012).

\bibitem{Y-Xray3}
E. Blackburn {\it et al}.,
Phys. Rev. Lett. {\bf 110}, 137004 (2013).


\bibitem{Bi-Xray1}
R. Comin {\it et al}.,
Science {\bf 343}, 390 (2014).

\bibitem{Bi-Xray2}
E.H. da Silva Neto {\it et al}.,
Science {\bf 343}, 393 (2014).

 \bibitem{Hg-Xray}
W. Tabis {\it et al}.,
Nat. Commun. {\bf 5}, 5875 (2014).

\bibitem{La-Xray}
M. H\"{u}cker {\it et al}.,
Phys. Rev. B {\bf 83}, 104506 (2011).

\bibitem{p-CDW}
R. Comin {\it et al}.,
arXiv:1402.5415.

\bibitem{DHLee-PNAS}
J. C. S. Davis and D.-H. Lee, 
Proc. Natl. Acad. Sci. USA, {\bf 110}, 17623 (2013).

\bibitem{Kivelson-NJP}
E. Berg, E. Fradkin, S. A. Kivelson, and J. M. Tranquada, 
New J. Phys. {\bf 11}, 115004 (2009).

\bibitem{Chubu}
Y. Wang and A.V. Chubukov, 
Phys. Rev. B {\bf 90}, 035149 (2014).

\bibitem{Sachdev}
M.A. Metlitski and S. Sachdev, New J. Phys. {\bf 12}, 105007 (2010);
S. Sachdev and R. La Placa, Phys. Rev. Lett. {\bf 111}, 027202 (2013).

\bibitem{Metzner}
C. Husemann and W. Metzner, Phys. Rev. B {\bf 86}, 085113 (2012);
T. Holder and W. Metzner, Phys. Rev. B {\bf 85}, 165130 (2012).

\bibitem{Bulut}
S. Bulut, W.A. Atkinson and A.P. Kampf,
Phys. Rev. B {\bf 88}, 155132 (2013).

\bibitem{Bianconi}
A. Bianconi, N. L. Saini, A. Lanzara, M. Missori, and T. Rossetti,
H. Oyanagi, H. Yamaguchi, K. Oka, and T. Ito,
Phys. Rev. Lett. {\bf 76}, 3412 (1996).

\bibitem{Onari-SCVC}
S. Onari and H. Kontani, 
Phys. Rev. Lett. {\bf 109}, 137001 (2012).

\bibitem{Ohno-SCVC}
Y. Ohno, M. Tsuchiizu, S. Onari, and H. Kontani, 
J. Phys. Soc. Jpn {\bf 82}, 013707 (2013).

\bibitem{Tsuchiizu}
M. Tsuchiizu, Y. Ohno, S. Onari, and H. Kontani, 
Phys. Rev. Lett. {\bf 111}, 057003 (2013).

\bibitem{Tsuchiizu2}
M. Tsuchiizu, Y. Yamakawa, Y. Ohno, S. Onari, and H. Kontani, 
arXiv:1405.2028.

\bibitem{Onari-SCVCS}
S. Onari, Y. Yamakawa, and H. Kontani, 
Phys. Rev. Lett. {\bf 112}, 187001 (2014).

\bibitem{Fernandes}
R.M. Fernandes {\it et al.},
Phys. Rev. Lett. {\bf 105}, 157003 (2010). 

\bibitem{OO-FeAs}
F. Kr\"{u}ger, S. Kumar, J. Zaanen, J. van den Brink, 
Phys. Rev. B {\bf 79}, 054504 (2009);
W. Lv, J. Wu, and P. Phillips, Phys. Rev. B {\bf 80}, 224506 (2009);
C.-C. Lee, W.-G. Yin, and W. Ku, Phys. Rev. Lett. {\bf 103}, 267001 (2009).

\bibitem{Kontani-Raman}
H. Kontani and Y. Yamakawa, 
Phys. Rev. Lett. {\bf 113}, 047001 (2014).

\bibitem{Held}
P. Hansmann, N. Parragh, A. Toschi, G. Sangiovanni, and K. Held, 
New J. Phys. {\bf 16}, 033009 (2014).

\bibitem{parameter-comment}
In the first principle study \cite{model-parameters},
interaction parameters are $(U_d,\ U_p,\ V)\approx(8,\ 3,\ 1)$ in eV.
In the RPA, however,
we have to put $U_d\sim4$ eV to avoid the SDW order,
since the self-energy correction, which describe the thermal and 
quantum fluctuations
that destroy the magnetic order, is absent in the RPA.
In fact, $\chi^s(\q)$ similar to Fig. \ref{fig:fig1} (c) 
is obtained for $U_d\gtrsim 8$ eV by using the fluctuation-exchange (FLEX) 
approximation thanks to the self-energy correction.
The result is not sensitive for $U_d\gtrsim8$ eV since the relation 
$\a_S<1$ is assured in the FLEX in 2D (Mermin-Wagner theorem); 
H. Kontani and M. Ohno, Phys. Rev. B {\bf 74}, 014406 (2006).

\bibitem{model-parameters}
M. S. Hybertsen, M. Schl\"{u}ter, and N. E. Christensen, 
Phys. Rev. B {\bf 39}, 9028 (1989).

\bibitem{Miyake}
K. Morita {\it et al}.,
J. Phys. Soc. Jpn {\bf 72}, 3164 (2003).

\bibitem{comment1}
In the LDA study, the $p$-orbital DOS, $N_p(0)$, 
is about 30\% of the total DOS, $N(0)=N_p(0)+N_d(0)$.
Although $N_p(0)/N(0)=0.46$ in the present $d$-$p$ model at $n=4.9$,
we can suppress the ratio $N_p(0)/N(0)\approx0.3$ by increasing $E_d-E_p$.
Even in this case, the CDW due to the VC is realized
by increasing $V$ just by $\sim30$\%.

\bibitem{suppl}
Supplemental Material.


\bibitem{doubleQ}
N. Harrison and S. E. Sebastian,
Phys. Rev. Lett. {\bf 106}, 226402 (2011).

\bibitem{smectic-theory}
H. Yao, D.-H. Lee, and S. Kivelson,
Phys. Rev. B {\bf 84}, 012507 (2011).

\bibitem{Kim}
Y.K. Kim {\it et al}.,
Science {\bf 345}, 187 (2014).

\bibitem{neutron-YBCO}
C. Stock, W. J. L. Buyers, R. Liang, D. Peets, Z. Tun, D. Bonn, 
W. N. Hardy, and R. J. Birgeneau,
Phys. Rev. B {\bf 69}, 014502 (2014).

\bibitem{Andersen}
S. Y. Savrasov and O. K. Andersen,
Phys. Rev. Lett. {\bf 77}, 4430 (1996).

\bibitem{Reznik}
D. Reznik, Advances in Condensed Matter Physics, Article ID 523549 (2010).

\bibitem{Dev}
S. Johnston {\it et al}.,
Phys. Rev. B {\bf 82}, 064513 (2010).


\bibitem{Kasahara}
S. Kasahara {\it et al}., 
Nature {\bf 486}, 382 (2012).


\end{thebibliography}
\end{document}